\documentclass[preprint,amsmath,amssymb,11pt,english,aps]{revtex4}
\usepackage{graphicx}
\usepackage{graphics}
\usepackage{amsmath}
\usepackage{dcolumn}
\usepackage{amssymb}
\usepackage{bm}
\usepackage[latin1]{inputenc}

\begin{document}
\title{Geometric Phase for Neutral Particle in the Presence of a Topological Defect}
\author{K. Bakke, J. R. Nascimento and C. Furtado}
\email{kbakke@fisica.ufpb.br,jroberto@fisica.ufpb.br,furtado@fisica.ufpb.br} 
\affiliation{Departamento de F\'{\i}sica, Universidade Federal da Para\'{\i}ba, Caixa Postal 5008, 58051-970, Jo\~o Pessoa, Pb, Brazil}

\begin{abstract}
In this paper we study the quantum dynamics of a neutral particle  in the presence of a topological defect. We investigate the appearance of a geometric phase in the relativistic quantum dynamics of neutral particle which possesses permanent magnetic
and electric dipole moments in the presence of an electromagnetic fields in this curved background. The nonrelativistic quantum dynamics  are investigated using the Foldy-Wouthuysen expansion. The gravitational Aharonov-Casher and He-Mckellar-Wilkens effects are investigated for a series of electric and magnetic fields configurations.  
\end{abstract}
\maketitle

\section{Introduction}

Topological defects are predicted in some unified theories of fundamental interactions. They may have been formed at phase transitions in the earliest history of the universe \cite{1}. Examples of such topological defects are the domain wall \cite{2}, the cosmic string \cite{2,3} and the global monopole \cite{4}. In particular, cosmic strings provide a bridge between the physical descriptions of  microscopic and macroscopic scales.

The appearance of topological phases in the quantum dynamics of a single particle moving freely in multiply connected spacetimes have been studied in a variety of physical systems. The prototype of this phase being the electromagnetic Aharonov-Bohm one \cite{5}, which appears as a phase factor in the wave function of an electron which moves around a magnetic flux line. The gravitational analog of this effect has also been studied in \cite{6,for,dow,anapr,bez}. Aharonov and Casher \cite{7} demonstrated that a magnetic dipole acquires a quantum phase when encircle a linear distribution of electric charge. A classical gravitational analog of Aharonov-Casher effect was investigated by Resnik \cite{11}. Also, He and McKellar \cite{8} and independently Wilkens \cite{9} have demonstrated that the quantum dynamics of an electric dipole in the presence of line of magnetic monopoles also exhibits a geometric quantum phase.
   
The gravitational  Aharonov-Bohm also was investigated by Mazur \cite{maz} for the relativistic quantum dynamics of particles in the presence of rotating cosmic string. In recent years, a series of authors was investigated geometric phase in the presence of gravitational fields. Cai and Papini \cite{papi,papi2} obtained a covariantly generalized form of the Berry phase and applied it to a situation involving weak gravitational field. Corichi and Pierri \cite{cori} studied a scalar quantum particle in the presence of rotating cosmic string and investigated the appearance of Berry geometric phase in this dynamics. Mostafazadeh \cite{ali} also considered the relativistic  Berry quantum phase in a series of problems involving scalar particles. In the paper \cite{assi} the gravitational Berry phase was applied to the quantum dynamics of scalar quantum particle in the presence of a chiral cosmic string. Shen  has carried out a series of studies concerning the Berry geometric phase in a curved space \cite{shen1,shen2,shen3}.
Recently, in \cite{mell} the non-relativistic quantum dynamics of  electric and magnetic dipole in the presence of a cosmic string was studied. This research was motivated by intention to investigate the quantum scattering and bound state of this dipole in the presence of an external electromagnetic field. 

In this paper we analyze the relativistic quantum dynamics of electric and magnetic dipole in the presence of topological defect with intention to investigate the influence of gravitational field in the geometric phase of electric and magnetic dipole in the presence of electric and magnetic fields. The relativistic geometric phase is obtained for a neutral particle. The Foldy-Wouthuysen approximation is used to investigate the gravitational Aharonov-Casher and the He-McKellar-Wilkens geometric phases.

The structure of the paper reads as follows. In the section II the geometric aspects of conical space is presented. In the section III, we investigate the relativistic quantum dynamics of electric and magnetic dipoles. In the section IV the Dirac equation in cosmic string space-time is analyzed. In the section V the Foldy-Wouthuysen approximation in the conical background is studied. In the section VI,  geometric quantum phase for investigated in the nonrelativistic quantum dynamic of neutral particle in conical background. Finally in the section VII the results are discussed.   

\section{The Cosmic string Background}\label{sec2}

In this section we develop the structure of the curved spacetime that we work out throughout this paper. We consider a  cosmic string space-time, where the line element is given by
\begin{eqnarray}
ds^{2}=-dt^{2}+d\rho^{2}+\eta^{2}\rho^{2}d\varphi^{2}+dz^{2}.
\label{2.1}
\end{eqnarray}
where  $\eta$ is called deficit angle and is defined as $\eta=1-4\mu $ where  $\mu$ is the linear mass density of the cosmic string. The azimuthal angle varies in the interval: $0\leq\varphi<2\pi$. The deficit angle can assume only values in which $\eta<1$ (unlike of this, in ~\cite{kat,furt}, it can assume values greater than 1, which correspond to an anti-conical space-time with negative curvature). This geometry possess a conical singularity represented by the following curvature tensor
\begin{eqnarray}
\label{curv}\label{curva}
R_{\rho,\varphi}^{\rho,\varphi}=\frac{1-\eta}{4\eta}\delta_{2}(\vec{r}),
\end{eqnarray}
where $\delta_{2}(\vec{r})$ is the two-dimensional delta function. This behavior of the curvature tensor is denominated conical singularity~\cite{staro}. The conical singularity gives rise to the curvature concentrated on the cosmic string axis, in all  other places the curvature is null.

It is convenient to construct a  frame which allows us to define the spinors in the curved spacetime. We can introduce the frame using a non-coordinate basis $\hat{\theta}^{a}=e^{a}_{\,\,\,\mu}\,dx^{\mu}$, which its components $e^{a}_{\,\,\,\mu}\left(x\right)$ satisfy the following relation \cite{bd,naka}
\begin{eqnarray}
g_{\mu\nu}\left(x\right)=e^{a}_{\,\,\,\mu}\left(x\right)\,e^{b}_{\,\,\,\nu}\left(x\right)\,\eta_{ab}.
\label{2.2}
\end{eqnarray}
The components of the non-coordinate basis $e^{a}_{\,\,\,\mu}\left(x\right)$ form \textit{tetrad or Vierbein}. The tetrad has a inverse define as $dx^{\mu}=e^{\mu}_{\,\,\,a}\,\hat{\theta}^{a}$, where 
\begin{eqnarray}
e^{a}_{\,\,\,\mu}\,e^{\mu}_{\,\,\,b}=\delta^{a}_{\,\,\,b}\,\,\,\,\,\,\,e^{\mu}_{\,\,\,a}\,e^{a}_{\,\,\,\nu}=\delta^{\mu}_{\,\,\,\nu}.
\label{2.3a}
\end{eqnarray}
For the metric corresponding to a cosmic string we choose the tetrad to be
\begin{eqnarray}
e^{a}_{\,\,\,\mu}=\left(
\begin{array}{cccc}
1 & 0 & 0 & 0 \\
0 & \cos\varphi & -\eta\rho\sin\varphi & 0 \\
0 & \sin\varphi & \eta\rho\cos\varphi & 0 \\
0 & 0 & 0 & 1 \\
\end{array}\right).
\label{2.4}
\end{eqnarray}
The tetrad inverse to the (\ref{2.4}) has the following form
\begin{eqnarray}
e^{\mu}_{\,\,\,a}=\left(
\begin{array}{cccc}
1 & 0 & 0 & 0 \\
0 & \cos\varphi & \sin\varphi & 0 \\
0 & -\frac{\sin\varphi}{\eta\rho} & \frac{\cos\varphi}{\eta\rho} & 0 \\
0 & 0 & 0 & 1 \\
\end{array}\right),
\label{2.3}
\end{eqnarray}
which yields the correct flat spacetime limit for $\eta=1$.
With the information about the choice of the frame, we can obtain the one-form connection $\omega^{a}_{\,\,\,b}=\omega_{\mu\,\,\,\,b}^{\,\,\,a}\,dx^{\mu}$ using  the Maurer-Cartan structure equation \cite{naka}
\begin{eqnarray}
d\hat{\Theta}^{a}+\omega^{a}_{\,\,\,b}\wedge\hat{\Theta}^{b}=0,
\label{2.5}
\end{eqnarray}
Hence, we obtain the following non-zero one-forms connections
\begin{eqnarray}
\omega_{\varphi\,\,\,\,\,2}^{\,\,\,\,1}=-\omega_{\varphi\,\,\,\,\,1}^{\,\,\,\,2}=1-\eta,
\label{2.6}
\end{eqnarray}

 Now we suggest that the core of cosmic string is charged and contain a linear distribution of magnetic  $\lambda_{m}$ and electric $\lambda_{e}$ charge. This configuration of charges produces a electric and magnetic fields given by\cite{mell} 

\begin{eqnarray}
\vec{E}=\frac{\lambda_{e}}{\eta\rho}\hat{\rho};\,\,\,\,\vec{B}=\frac{\lambda_{m}}{\eta\rho}\hat{\rho},
\label{2.6.1}
\end{eqnarray}
Notice that in the limit where $\eta\longrightarrow1$ we obtain well known results to electric (magnetic) fields produced by a linear density of electric (magnetic) charge in the Minkowisky space-time.

\section{Relativistic Quantum Dynamics}
 
In this section we consider the quantum dynamics of a neutral spin-half particle with nonzero magnetic and electric dipole moments.
We analyze the Dirac equation in a curved space-time in the presence of an electric and magnetic fields. The Dirac equation with a non-minimal coupling of the spinor to the electromagnetic field embedded in a classical gravitational field is given by
\begin{eqnarray}
i\gamma^{\mu}\,\nabla_{\mu}\psi+\frac{\mu}{2}\Sigma^{\mu\nu}\,F_{\mu\nu}\psi-i\frac{d}{2}\,\Sigma^{\mu\nu}\,\gamma^{5}\,F_{\mu\nu}-m\psi=0,
\label{1.1}
\end{eqnarray}
where $\mu$ is the magnetic dipole moment, $d$ is the electric dipole moment, and
\begin{eqnarray}
\nabla_{\mu}=\partial_{\mu}+\Gamma_{\mu}
\label{3.1}
\end{eqnarray}
is the components of covariant derivative and $\Gamma_{\mu}$ is the spinor connection \cite{bd} which is given by 
\begin{eqnarray}
\Gamma_{\mu}&=&\frac{1}{8}\omega_{\mu ab}\left(x\right)\left[\gamma^{a},\gamma^{b}\right]\\
&=&\frac{1}{8}\,e_{a\nu}\,\nabla_{\mu}\,e^{\nu}_{\,\,\,b}\left[\gamma^{a},\gamma^{b}\right].
\label{3.2}
\end{eqnarray} and  
\begin{eqnarray}
F_{\mu\nu}=\nabla_{\mu}A_{\nu}-\nabla_{\nu}A_{\mu}=\left\{\vec{E},\vec{B}\right\}
\end{eqnarray}
with $F_{0\alpha}=-F_{\alpha 0}=E_{\alpha}$, $F_{\alpha\beta}=-F_{\beta\alpha}=-\epsilon_{\alpha\beta\gamma}\,B^{\gamma}$ and $(\alpha,\beta,\gamma=\rho,\varphi,z)$ are the spatial indices of the spacetime. The matrices $\gamma^{\mu}$ are generalized Dirac matrices given in terms of the flat spacetime ones $\gamma^{a}$  by the relation $\gamma^{\mu}=e^{\mu}_{\,\,\,a}\gamma^{a}$.   We rewrite the Dirac equation (\ref{1.1}) in terms of vierbeins in the following form
\begin{eqnarray}
i\gamma^{a}\,e^{\mu}_{\,\,\,a}\,\nabla_{\mu}\psi+\frac{1}{2}\mu\,F_{\mu\nu}\,e^{\mu}_{\,\,\,a}\,e^{\nu}_{\,\,\,b}\,\Sigma^{ab}\psi
-\frac{i}{2}d\,\Sigma^{ab}\,e^{\mu}_{\,\,\,a}\,e^{\nu}_{\,\,\,b}\gamma^{5}\,F_{\mu\nu}-m\psi=0,
\label{1.2}
\end{eqnarray} 
with $\Sigma^{ab}=\frac{i}{2}\left[\gamma^{a},\gamma^{b}\right]$ and $(a,b,c=0,1,2,3)$ are the indices which indicates the local reference frame. The $\gamma^{a}$ matrices are the Dirac matrices in the flat spacetime, \textit{i.e.},
\begin{eqnarray}
\gamma^{0}=\hat{\beta}=\left(
\begin{array}{cc}
1 & 0 \\
0 & -1 \\
\end{array}\right),\,\,\,\,\,\,
\gamma^{i}=\hat{\beta}\,\alpha^{i}=\left(
\begin{array}{cc}
 0 & \sigma^{i} \\
-\sigma^{i} & 0 \\
\end{array}\right),
\end{eqnarray}
with $\sigma^{i}$ are the Pauli matrices satisfying the relation $\left(\sigma^{i}\,\sigma^{j}+\sigma^{j}\,\sigma^{i}\right)=-2\,\eta^{ij}$, and $\eta^{ab}=diag(-1, 1, 1, 1)$ is the Minkowsky tensor and the index $(i,j,k=1,2,3)$ are the spatial index of the local reference frame. The $\gamma^{5}$ matrix is defined as
\begin{eqnarray}
\gamma^{5}&=&-\frac{i}{24}\,\epsilon_{\mu\nu\eta\lambda}\,\gamma^{\mu}\,\gamma^{\nu}\,\gamma^{\eta}\,\gamma^{\lambda}=
i\,\gamma^{0}\,\gamma^{1}\,\gamma^{2}\,\gamma^{3}=-\gamma_{5} =\left(
\begin{array}{cc}
0 & 1 \\
1 & 0 \\
\end{array}\right),
\end{eqnarray}
and, finally, we write  $\vec{\Sigma}$ as
\begin{eqnarray}
\vec{\Sigma}=\left(
\begin{array}{cc}
\vec{\sigma} & 0 \\
0 & \vec{\sigma} \\
\end{array}\right),
\end{eqnarray}
whose components are defined in the local reference frame. 

\section{Dirac equation in Cosmic string Background}
$ $ 
The Dirac equation that describes a spin-$1/2$  neutral particle with non-zero magnetic and electric dipole moments moving in an external electromagnetic field is given by the expressions (\ref{1.1}) or (\ref{1.2}).
Using the expression (\ref{2.6}), the spinorial connection has  only the following non-zero component
\begin{eqnarray}
\Gamma_{\varphi}&=&\frac{1}{4}\left(1-\eta\right)\left[\gamma^{1},\gamma^{2}\right]=
-\frac{i}{2}\left(1-\eta\right)\,\Sigma^{3}.
\label{3.2a}
\end{eqnarray}

Thus, the Dirac equation in curved spacetime (\ref{1.2}) has the form 
\begin{eqnarray}
&&i\,\gamma^{t}\,\frac{\partial\psi}{\partial t}+i\,\gamma^{\rho}\left(\partial_{\rho}+\frac{1}{2}\frac{\left(1-\eta\right)}{\eta\rho}+\mu\,E_{\rho}-d\,B_{\rho}\right)\psi+\nonumber\\&+&
i\frac{\gamma^{\varphi}}{\eta\rho}\,\frac{\partial\psi}{\partial\varphi}+i\,\gamma^{z}\,\frac{\partial\psi}{\partial z}-\mu\,\vec{\Sigma}\cdot\vec{B}\,\psi-d\,\vec{\Sigma}\cdot\vec{E}\,\psi -m\psi=0.
\label{3.7a}
\end{eqnarray}
In this background the matrices $\gamma^{\mu}=e^{\mu}_{\,\,\,a}\,\gamma^{a}$ are given by
\begin{eqnarray}
\gamma^{t}&=&e^{t}_{\,\,\,a}\,\gamma^{a}=\gamma^{0};\,\,\,\,\gamma^{z}=e^{z}_{\,\,\,a}\,\gamma^{a}=\gamma^{3};\\
\gamma^{r}&=&e^{r}_{\,\,\,a}\,\gamma^{a}=\cos\varphi\,\gamma^{1}+\sin\varphi\,\gamma^{2};\\
\gamma^{\varphi}&=&e^{\varphi}_{\,\,\,a}\,\gamma^{a}=-\sin\varphi\,\gamma^{1}+\cos\varphi\,\gamma^{2}.
\label{3.7b}
\end{eqnarray}

Now, let us discuss the relativistic geometric phase in this dynamics. We consider that the spinor $\psi$ can be written in the following form
\begin{equation}\label{phased}
 \psi=e^{i\phi}\psi_{0}
\end{equation}
where $\psi_{0}$ is phase and is the solution of Dirac equation in the absence of fields and $\phi$ is a phase. Substituting  Eq.(\ref{phased}) in  Eq. (\ref{3.7a}) we obtain the following phase
\begin{equation}\label{relatpha}
\phi =\oint \left(\frac{1}{2}\left(1-\eta\right)\,\Sigma^{3}-\mu \,\left(\vec{\Sigma}\times\vec{E}\right)_{\varphi}+ d \,\left(\vec{\Sigma}\times\vec{B}\right)_{\varphi}\right)\,d\varphi,
\end{equation}
Note that this phase has three contribution, the first contribution is generated by the conical geometry of cosmic string background. The other two contributions are generated by the dipole interaction and depend of the field configuration. Note that the first term in Eq. (\ref{relatpha}) can be written in the following form
\begin{equation}\label{pap}
 \phi_{p}= \frac{1}{4}\oint R_{\mu \nu \delta \lambda} J^{ \delta \lambda}d\tau^{\mu \nu}
\end{equation}
where $J^{ \delta \lambda}= L^{ \delta \lambda} + \Sigma^{ \delta \lambda}$ is the total angular momentum of the particle.
Substituting here the curvature tensor given by   Eq.(\ref{curva}), we find that the expression (\ref{pap}) takes the following form:
\begin{equation}
\phi_{p}=\oint \frac{1}{2}\left(1-\eta\right)\,\Sigma^{3}  d\varphi
\end{equation}
This phase is the relativistic  Berry's geometric phase proposed by  Cai and Papini \cite{papi,papi2} for spin-$1/2$ particle in a curved background using a weak field approximation. In our study, we do not use this approximation, and obtain that the phase found in \cite{papi,papi2} is generic. 

Other two terms  in (\ref{relatpha}) are the contributions due to the magnetic  and electric dipole moments to relativistic Anandan geometric phase \cite{12,13} in the presence of cosmic string  which is given by
\begin{equation}\label{relatpha1}
\phi =\oint \left(- \mu \,\left(\vec{\Sigma}\times\vec{E}\right)_{\varphi}+d \,\left(\vec{\Sigma}\times\vec{B}\right)_{\varphi}\right)\,d\varphi,
\end{equation}
In the limit that $\eta\longrightarrow 1$ we obtain flat spacetime results for the Anandan geometric phase.

\section{Nonrelativistic Limit}

In this section we investigate the nonrelativistic limit for a spinor particle non-minimally coupled to electromagnetic fields embedded in a classical gravitational field. We will use the Foldy-Wouthuysen method \cite{fw,silenko1}.
First, let us rewrite the Dirac equation (\ref{1.1}) in the following form:
\begin{eqnarray}
i\frac{\partial\psi}{\partial t}=H\psi.
\label{3.3}
\end{eqnarray}
After some manipulation, we arrive at the following equation
\begin{eqnarray}
i\frac{\partial\psi}{\partial t}&=&m\hat{\beta}\psi-\vec{\alpha}\cdot\vec{p}\psi-i\,\vec{\alpha}\cdot\vec{\xi}\psi\nonumber\\
&+&\mu\hat{\beta}\left(-i\vec{\alpha}\cdot\vec{E}+\vec{B}\cdot\vec{\Sigma}\right)\psi\nonumber\\
&+&d\hat{\beta}\left(i\vec{\alpha}\cdot\vec{B}-\vec{E}\cdot\vec{\Sigma}\right)\psi,
\label{3.4}
\end{eqnarray}
where we  has defined the following  terms $p_{j}=-i\,e^{\alpha}_{\,\,\,j}\,\partial_{\alpha}$, $E_{j}=e^{\alpha}_{\,\,\,j}\,E_{\alpha}$, $B_{j}=e^{\alpha}_{\,\,\,j}\,B_{\alpha}
$ and $\xi_{j}= e^{\varphi}_{\,\,\,j}\,\Gamma_{\varphi}$, which in a cosmic string background  is given by
\begin{eqnarray}
 \xi_{j}=-\frac{i}{2}\left(1-\eta\right)\,\Sigma^{3}\,e^{\varphi}_{\,\,\,j},
\label{3.5b}
\end{eqnarray}
Thus, we can write the Dirac equation (\ref{3.4}) in the form
\begin{eqnarray}
i\frac{\partial\psi}{\partial t}=m\hat{\beta}\psi+\vec{\alpha}\cdot\vec{\pi}\psi+d\vec{E}\cdot\vec{\Sigma}\psi
+\mu\hat{\beta}\vec{B}\cdot\vec{\Sigma}\psi,
\label{3.6}
\end{eqnarray} 
where the operator $\vec{\pi}$, in the local reference frame, is defined as 
\begin{eqnarray}
\vec{\pi}=\vec{p}-i\mu\,\hat{\beta}\,\vec{E}+id\,\hat{\beta}\,\vec{B}-i\vec{\xi},
\label{3.7}
\end{eqnarray}
and the three first terms has the same form as defined by \cite{fur2} in the flat spacetime. The last term of (\ref{3.7}) arises due the topology of the spacetime.

We investigate the non-relativistic limit of the Dirac equation using the Foldy-Wouthuysen approximation \cite{fw}. In this approximation, the Hamiltonian of the system is written as a following linear combination
\begin{eqnarray}
H=\hat{\beta}\,m+\hat{O}+\hat{\epsilon},
\label{4.1}
\end{eqnarray}
where the operators $\hat{O}$ and $\hat{\epsilon}$ should be Hermitian ones and must satisfy the relations
\begin{eqnarray}
\begin{array}{c}
\hat{O}\,\hat{\beta}+\hat{\beta}\,\hat{O}=0,\\
\hat{\epsilon}\,\hat{\beta}-\hat{\beta}\,\hat{\epsilon}=0.\\
\end{array}
\label{4.2}
\end{eqnarray}

The final result obtained in this approximation permits us to expand the Hamiltonian $H$ and consider the terms up to the order of $m^{-1}$. So, we have
\begin{eqnarray}
H'''=\hat{\beta}\,m+\frac{\hat{\beta}}{2m}\hat{O}^{2}+\hat{\epsilon}.
\label{4.3}
\end{eqnarray}  
Using the expression (\ref{3.6}) we have that 
\begin{eqnarray}
\hat{O}&=&\vec{\alpha}\cdot\vec{\pi},\\
\hat{\epsilon}&=&\mu\,\hat{\beta}\,\vec{B}\cdot\vec{\Sigma}+d\,\hat{\beta}\,\vec{E}\cdot\vec{\Sigma},
\label{4.4}
\end{eqnarray}
and the expression for the Hamiltonian (\ref{4.3}) becomes
\begin{eqnarray}
H'''&=&\hat{\beta}\,m+\frac{\hat{\beta}}{2m}\left(\vec{p}+\vec{\Xi}\right)^{2}-\frac{\mu^{2}\,E^{2}}{2m}-\frac{d^{2}\,B^{2}}{2m}\nonumber\\
&+&\frac{\mu}{2m}\vec{\nabla}\cdot\vec{E}-\frac{d}{2m}\vec{\nabla}\cdot\vec{B}+d\,\hat{\beta}\,\vec{\Sigma}\cdot\vec{E}\nonumber\\
&+&\mu\,\hat{\beta}\,\vec{\Sigma}\cdot\vec{B},
\label{4.5}
\end{eqnarray}
where $\vec{\nabla}$ refers to the gradient in the spacetime indices and we introduce the vector, which components are
\begin{eqnarray}
\Xi_{j}=\mu\,\hat{\beta}\,(\vec{\Sigma}\times\vec{E})_{j}-d\,\hat{\beta}\,(\vec{\Sigma}\times\vec{B})_{j}+\frac{1}{2}\left(1-\eta\right)\Sigma^{3}\,e^{\varphi}_{\,\,\,j},
\label{4.6}
\end{eqnarray}
which is well defined in the local reference frame.

The Hamiltonian given in (\ref{4.5}) describes the behavior of the electric and magnetic dipoles in the external electric and magnetic fields with the presence of a  topological defect. The influence  of the topological defect (\ref{3.1}) becomes clear due to the third term  in the expression in Eq.(\ref{4.6}).  We can see that if we consider the limit $\eta\rightarrow 1$, \textit{i.e.}, the absence  of the topological defect, we arrive at the configuration obtained in \cite{fur2} for the flat spacetime. The effects arisen due to the configuration of the dipoles in the presence of  a topological defect with the influence of external electric and magnetic fields will be discussed in the next section.

\section{Nonrelativistic Geometric Quantum Phases}

Now let us study the effects of interference of the neutral particles in the presence of the topological defect and the influence of the external electric and magnetic  fields. We consider the terms which contribute to the appearance of the geometric phase in the wave function. The non-relativistic Hamiltonian describing a neutral particle that possesses constant electric and magnetic dipole moments, in the presence of electric field embedded in a classical gravitational field, can be written in a following way
\begin{eqnarray}\label{nrham}
H=-\frac{1}{2m}\left( \vec{\nabla}- i\vec{\Xi}\right)^{2} + \Xi_{0}
\end{eqnarray} 
where $\Xi_{i}=\mu \,\hat{\beta}\,(\vec{\Sigma}\times\vec{E})_{i}- d\,\hat{\beta}\,(\vec{\Sigma}\times\vec{B})_{i}+\frac{1}{2}\left(1-\eta\right)\Sigma^{3}\,e^{\varphi}_{\,\,\,i},$  and $\Xi_{0}$ is given by
\begin{equation}\label{b0}
 \Xi_{0}=-\frac{\mu^{2}\,E^{2}}{2m}-\frac{d^{2}\,B^{2}}{2m}
+\frac{\mu}{2m}\vec{\nabla}\cdot\vec{E}-\frac{d}{2m}\vec{\nabla}\cdot\vec{B}+d\,\hat{\beta}\,\vec{\Sigma}\cdot\vec{E}
+\mu\,\hat{\beta}\,\vec{\Sigma}\cdot\vec{B},
\end{equation}
Notice that the expression (\ref{nrham}) is similar to Hamiltonian of quantum particle is minimally coupled to a non-Abelian gauge field $\Xi_{\mu}$. We can investigate the geometric phase of this system in the fields configuration given by (\ref{2.6.1}) and consider the dipoles oriented along $z$-direction. The last four terms in (\ref{b0}) give zero contribution to the geometrical phase for the field-dipole configuration adopted here.  The terms proportional to $E^{2}$ and $B^{2}$ are local terms and do not contribute for geometric phase \cite{12,13,fur2}. So that the only terms that contribute to geometric phase in (\ref{nrham}).
We have following equation
\begin{eqnarray}
-\frac{1}{2m}\left(\vec{\nabla}-i\mu\,\hat{\beta}\,\vec{\Sigma}\times\vec{E}+ id\,\hat{\beta}\,\vec{\Sigma}\times\vec{B}-i\frac{1}{2}\left(1-\eta\right)\Sigma^{3}\,e^{\varphi}\right)^{2}\,\Psi-\frac{\mu^{2}\,E^{2}}{2m}\,\Psi-\frac{d^{2}\,B^{2}}{2m}\,\Psi=E\,\Psi.
\label{5.1}
\end{eqnarray}   

The quantum phase can be obtained if we consider the ansatz:
\begin{eqnarray}
\Psi=e^{i\Phi}\,\psi,
\label{5.2}
\end{eqnarray}
where $\psi$ is the solution of the equation
\begin{eqnarray}
-\frac{1}{2m}\,\nabla^{2}\psi-\frac{\mu^{2}\,E^{2}}{2m}\,\psi-\frac{d^{2}\,B^{2}}{2m}\,\psi=E\,\psi.
\label{5.3}
\end{eqnarray}
Now, we analyze the quantum geometric phase when we consider the charge densities concentrated on the symmetry axis. In that way the fields are cylindrically symmetric and are given by Eq. (\ref{2.6.1}). Hence, taking into account the local reference frame and that the charges are concentrated on the symmetry axis of the topological defect, the quantum geometric phase of this system is given by
\begin{eqnarray}
\Phi&=&\oint\,\Xi_{\mu}\,dx^{\mu}=\oint\,\Xi_{i}\,e^{i}_{\,\,\,\mu}\,dx^{\mu}=\int^{2\pi}_{0}\,\Xi_{i}\,e^{i}_{\,\,\,\varphi}\,\,d\varphi\nonumber\\
&=&\left(1-\eta\right)\,\pi\,\sigma^{3}+\left(\mu\,\lambda_{e}-d\,\lambda_{m}\right)\,2\pi\,\sigma^{3}.
\label{5.5}
\end{eqnarray}
Here we have considered only two-component spinor fields.The geometric phase (\ref{5.5}) is a generalization of the Anandan quantum phase in the presence of a cosmic string. The contribution due to the defect can be seen from the first term  in Eq. (\ref{5.5}). If we take $\eta\rightarrow 1$, we recuperate the results obtained in \cite{fur1} in the absence of the topological defect. If we consider that the particle does not possess an electric dipole moment, that is, apply the limit  $d=0$ in (\ref{5.5}), we obtain the analog of the Aharonov-Casher effect in the presence of a topological defect. The quantum phase in this case becomes  
\begin{eqnarray}
\Phi_{AC}&=&i\,\oint\,e^{j}_{\,\,\,\varphi}\,\xi_{j}\,d\varphi+\mu\,\hat{\beta}\,\oint\left(\vec{\Sigma}\times\vec{E}\right)_{\varphi}\,d\varphi \nonumber\\
&=&\left(1-\eta\right)\,\pi\,\sigma^{3}+2\pi\,\mu\,\lambda_{e}\,\sigma^{3}
\label{5.6}
\end{eqnarray}
Note that we have a topological contribution to Aharonov-Casher effect due to the defect, and this term  gives topological nonvanishing contribution to the geometric phase \cite{bud}. Note that in the limit  $\eta\rightarrow 1$ we obtain the well known  Aharonov-Casher geometric phase. 

Our next step is consider the limit where magnetic moment of the particle is zero, $\mu=0$  in Eq.(\ref{5.5}), and  we obtain the analog of the He-McKellar-Wilkens effect in the presence of a topological defect. The quantum phase in this case becomes 
\begin{eqnarray}
\Phi_{HMW}&=&i\,\oint\,e^{j}_{\,\,\,\varphi}\,\xi_{j}\,d\varphi-d\,\hat{\beta}\,\oint\left(\vec{\Sigma}\times\vec{B}\right)_{\varphi}\,d\varphi \nonumber\\
&=&\left(1-\eta\right)\,\pi\,\sigma^{3}- \,2\pi\,d\,\lambda_{m}\,\sigma^{3}.
\label{5.7}
\end{eqnarray}
Both of the results obtained above demonstrate the influence of the topological defect to the geometric phase acquired by the wave function in the dynamics of the neutral particle in the presence of a cosmic string. If we consider the absence of the external field and only the presence of the defect, we will obtain a quantum phase that depends only  of the topological defect (\ref{3.1}). 
\begin{eqnarray}
\Phi=\left(1-\eta\right)\,\pi\,\sigma^{3}.
\label{5.8} 
\end{eqnarray}
The same contribution was obtained in  \cite{fur3}, when the holonomy matrix is found for spinor in a continuum model for a graphene layer with a topological defect. In that way, the general result given in the expression (\ref{5.5}) show us the geometric phase acquired in the dynamics of the neutral particle with a permanent electric and magnetic moments of dipoles is influenced by the presence of a topological defect.

\section{Conclusion}

We studied the influence  of the topological  defect in the geometric phases of dipoles in relativistic dynamics. We found a new contribution to geometric phase due the presence of topological defect. This contribution is a nondispersive topological \cite{pes2} contribution to total geometrical phase acquired by the neutral particle. This contribution is of gravitational origin \cite{papi,papi2} due curvature introduced by defect in the space-time. We have investigated the nonrelativistic quantum phase in the present paper using the Foldy-Wouthuysen approximation to obtain the nonrelativistic Hamiltonian. We saw that  the  topological defect introduce  a new gravitational contribution to Anandan's geometric phase. We can see that when $\eta\rightarrow 1$, we recuperate the same phase obtained in \cite{12,13} for a flat space  case in absence of a topological defect and also the phase  given in expressions (\ref{5.5}), (\ref{5.6}) and (\ref{5.7}) becomes the same given in \cite{fur1}. Note that the presence of a topological defect introduce a new term of the topological nature in  the Aharonov-Casher and He-McKellar-Wilkens geometric phase. We claim that this results can be interesting to investigate geometric phase in context of topological defect in condensed matter\cite{epl1,epl2}, where appear a class of linear topological defects of same nature of cosmic strings\cite{kat,furt}

\acknowledgments{Authors are gratfull to A. Yu. Petrov for some criticism on the manuscript. This work was partially supported by PRONEX/FAPESQ-PB, FINEP, CNPq and CAPES/PROCAD)}


\begin{thebibliography}{99}

\bibitem{1} T. W. B. Kibble, J. Phys. A {\bf 19}, 1387 (1976).

\bibitem{2} A. Vilenkin, Phys. Rep. {\bf 121}, 263 (1985).

\bibitem{3} A. Vilenkin, Phys. Lett. B {\bf 133}, 177 (1983); W. A. Hiscock, Phys. Rev. A {\bf 31}, 3288 (1985); B. Linet, Gen. Rel. Grav. {\bf 17}, 1109 (1985). 

\bibitem{4} M. Barriola and A. Vilenkin, Phys. Rev. Lett. {\bf 63}, 341 (1989).

\bibitem{5} Y. Aharonov and D. Bohm, Phys. Rev.{\bf 115}, 485 (1959).

\bibitem{6} L. H. Ford, A. Vilenkin, J. Physics A: Math. Gen. {\bf 14}, 2353 (1981); J. S. Dowker, Nuovo Cim. {\bf 52}, 129 (1967); J. Anandan, Phys. Lett. A {\bf 195}, 284 (1994); V. Bezerra, Phys. Rev. D {\bf 35}, 2031 (1987).

\bibitem{for} L. H. Ford and A. Vilenkin, J. Phys. {\bf A 14}, 2353 (1981) 

\bibitem{dow} J. S. Dowker, Nuovo Cim. {\bf 52}, 129 (1967).

\bibitem{anapr}J. Anandan, Phys. Lett. {\bf  A 195}, 284 (1994). 

\bibitem{bez} V. B. Bezerra, Phys. Rev. {\bf D 35}, 2031   (1987).

\bibitem{7} Y, Aharonov and A, Casher, Phys. Rev. Lett.{\bf 53}, 319 (1984).

\bibitem{11} B. Resnik, Phys. Rev. D {\bf 51} 3108 (1995).

\bibitem{8} X.-G. He, B. H. J. McKllar, Phys. Rev. A {\bf47}, 3424 (1983).

\bibitem{9} M. Wilkens, Phys. Rev. Lett.{\bf 72}, 5 (1994).
 
\bibitem{maz}P. O. Mazur, Phys. Rev. Lett. {\bf 57} 929,(1986).

\bibitem{papi}Y. Q. Cai and G. Papini, Mod. Phys. Lett. {\bf A 4}, 1143 (1989) .

\bibitem{papi2}Y. Q. Cai and G. Papini, Class. Quantum Grav. {\bf 7}, 269 (1990) .

\bibitem{cori} A. Corichi and M. Pierri, Phys. Rev. {\bf D 51}, 5870 1995 

\bibitem{ali}A. Mostafazadeh, J. Phys. {\bf A 31}, 7829 (1998).

\bibitem{assi}J. G. de Assis, C. Furtado, and V. B. Bezerra, Phys. Rev.{\bf  D 62}, 045003 (2000).

\bibitem{shen1}Jian Qi Shen, Physica Scripta {\bf 73}, 79 (2006).

\bibitem{shen2}J. Q. Shen,  Eur. Phys. J. D {\bf 30}, 259 (2004).

\bibitem{shen3}Jian Qi Shen, J. Opt. B {\bf  6}, L13 (2004).

\bibitem{mell}E. R. Bezerra de Mello, JHEP {\bf 04} 06  016 (2004).

\bibitem{10} H. Wei, R. Han and X. Wei, Phys. Rev. Lett. 75, 2071 (1995).

\bibitem{kat} M. O. Katanaev   and   I. V.  Volovich,  Ann. Phys. (NY) {\bf 216}, 1 (1992).

\bibitem{furt}C. Furtado and F. Moraes, Phys. Lett. {\bf A188}, 392 (1994).

\bibitem{staro} D.D. Sokolov and A.A. Starobinskii, Sov. Phys. Dokl. {\bf 22}, 312 (1977).

\bibitem{12} J. Anandan, Phys. Rev. Lett {\bf 85}, 1354 (2000).

\bibitem{13} J. Anandan, Phys. Lett. A  {\bf 138},347 (1989).

\bibitem{bd} Birrel and Davies, \textit{Quantum Fields in Curved Space}, Cambridge Univ. Press, Cambridge, UK, 1982.

\bibitem{naka} M. Nakahara, \textit{Geometry, Topology and Physics}, (Institute of Physics Puclishing Bristol 1998). 

\bibitem{fur1} C. Furtado and C. A. de Lima Ribeiro, Phys. Rev. A {\bf 69}, 064104 (2004).

\bibitem{bm1} B. Mashhonn, Phys. Rev. D {\bf 8}, 4297 (1973).

\bibitem{fur2} E. Passos, L. R. Ribeiro, C. Furtado and J. R. Nascimento, Phys. Rev. A {\bf 76}, 012113 (2007).
\bibitem{bud}G. Badurek, H. Weinfruter, R. G\"{a}hler, A. Kollmar, S. Wehinger and A. Zeilinger, Phys. Rev. Lett. {\bf 71}, 307 (1993)
\bibitem{fur3} C. Furtado, F. Moraes and A. M. de M. Carvalho, cond-mat 0601391.

\bibitem{fw} L.L. Foldy and S. A. Wouthuysen, Phys. Rev.{\bf 78}, 29 (1950).
\bibitem{silenko1} A. J. Silenko, Russ. Phys. J. {\bf 48},788, (2005).
\bibitem{pes2}M.  Peshkin  and H. J. Lipkin, Phys. Rev. Lett. {\bf 74}, 2847 (1995).
\bibitem{epl1}C. Furtado, V. B. Bezerra and F. Moraes, Europhys. Lett. {\bf 52}, 1 (2000).
\bibitem{epl2} C. A. de Lima Ribeiro, C Furtado and F. Moraes, Europhys. Lett. {\bf 62} 306 (2003).
\end{thebibliography}
\end{document}